\documentclass{ws-procs975x65}

\def\lsim{\mathop{\hbox{${\lower3.8pt\hbox{$<$}}\atop{\raise0.2pt\hbox{$\sim$}}
$}}} 
\def\gsim{\mathop{\hbox{${\lower3.8pt\hbox{$>$}}\atop{\raise0.2pt\hbox{$
\sim$}}$}}}

\usepackage{color}
\definecolor{MyGreen}{rgb}{0.21,0.64,0.18}
\definecolor{MyCyan}{rgb}{0.0,0.6,0.6}
\definecolor{MyDarkRed}{rgb}{0.71,0.14,0.07}

\newcommand{\CR}{{\cal R}}

\newcommand{\initial}[1]{{#1_{\rm \bf i}}}

\usepackage[dvips,breaklinks]{hyperref}
\hypersetup{colorlinks,urlcolor=black,citecolor=black,linkcolor=black,filecolor=black}
\usepackage{breakurl}

\begin{document}

\markboth{Al Roumi \& Buchert}
         {Relativistic structure formation models and gravitoelectromagnetism}

\title{Relativistic structure formation models and gravitoelectromagnetism}

\author{Fosca Al Roumi and Thomas Buchert} 

\address{Universit\'e de Lyon, Observatoire de Lyon, 
Centre de Recherche Astrophysique de Lyon,\\ CNRS UMR 5574: Universit\'e Lyon~1 and \'Ecole Normale Sup\'erieure de Lyon,\\
Site Jacques Monod, 46 all\'ee d'Italie, F--69364 Lyon cedex 07}



\begin{abstract}
In the framework of Lagrangian perturbation theory in general relativity we discuss the possibility to split the Einstein equations, written in terms of spatial Cartan coframes within a $3+1$ foliation of spacetime, into gravitoelectric and gravitomagnetic parts. While the former reproduces the full hierarchy of the Newtonian perturbation solutions, the latter 
contains non--Newtonian aspects like gravitational waves. This split can be understood and made unique through the Hodge decomposition of Cartan coframe fields. 
\end{abstract}



\section{Introduction}

We discuss an intrinsic formulation of relativistic perturbation theory.\cite{rza1} In this novel approach the perturbation fields evolve on the physical manifold rather than on  a pre--defined background spacetime, as is commonly assumed in the framework of the standard perturbation theory.\par
We first motivate our formalism by paraphrasing the Lagrangian perturbation approach in Newtonian theory. The relativistic approach leads to a closed system for
a single perturbation variable, summarized as the {\em Lagrange--Einstein system}. It is obtained from a $3+1$ decomposition of the Einstein equations and formulated in terms of spatial Cartan coframes. We will then employ the gravitoelectromagnetic analogy that allows us to split the {\em Lagrange--Einstein system} into gravitoelectric and gravitomagnetic parts.
We will explain how the gravitoelectric hierarchy of perturbation solutions\cite{rza3} can be obtained from Newtonian perturbation theory.\cite{ehlersbuchert}\par
Assuming a closed topological space form for the spatial hypersurfaces, we employ the Hodge decomposition (that replaces the Helmholtz decomposition in standard perturbation theory) to fully understand the physical content of the gravitoelectric and gravitomagnetic solutions, exemplified to first order: the gravitoelectric part will turn out to be the integrable part of the Cartan deformations, while the nonintegrable, gravitomagnetic part contains the gravitational waves.

\section{Intrinsic Lagrangian approach to structure formation}

\subsection{Newtonian description of structure formation}
We consider a dust continuum (i.e., without pressure and velocity dispersion), governed by the Euler--Newton system of equations:
 \begin{align} 
 \label{ENS}
\frac{\partial \mathbf{v}}{ \partial t} + \mathbf{v} \cdot \boldsymbol{\nabla}\mathbf{v} =  \mathbf{g} \qquad;\qquad \frac{\partial \varrho}{ \partial t}  +  \boldsymbol{\nabla}\cdot ( \varrho \mathbf{v}) =  0\:\:,\\
\boldsymbol{\nabla}\times  \mathbf{g} = \mathbf{0}\qquad;\qquad\boldsymbol{\nabla}\cdot  \mathbf{g} =  \Lambda - 4 \pi G \varrho\:\:,
\end{align}
where $ \mathbf{v}$ is the Eulerian velocity, $\mathbf{g} $ the gravitational field strengh, and $\varrho$ the density.
We represent by $\mathbf{f}({\small{\bullet}},t)$ the diffeomorphism\footnote{The description in terms of the trajectory function $\mathbf{f}$ is only possible before shell-crossing. Once caustics have formed, $\mathbf{f}({\small{\bullet}},t)$ is no longer a diffeomorphism and this description breaks down.} between the Lagrangian spatial coordinates $ \mathbf{X} $  which label fluid elements, and the Eulerian ones $ \mathbf{x} $, which are now considered as values of a position field $ \mathbf{f} $ of these elements at time $t$:
\begin{equation}
\mathbf{f}({\small{\bullet}},t): \mathbb{R}^3 \longrightarrow \mathbb{R}^3 \;\;;\;\;
             \mathbf{X} \longmapsto  \mathbf{x}=  \mathbf{f}( \mathbf{X},t) \quad \mbox{and} \quad \mathbf{X} := \mathbf{f} (\mathbf{x},t_i)\;.
\label{trajectory}
\end{equation}
We define the determinant of the Jacobian matrix $(f_{i \vert j})$ by:
\begin{equation}
J := \det \left(\frac{\partial f_i}{ \partial X_j} (\mathbf{X},t)\right) = \frac16 \epsilon_{ijk}  \epsilon^{klm} f^i_{\ \vert k}f^j_{ \ \vert l} f^k_{\ \vert m}\:\:,
\end{equation}
where we have denoted by a vertical slash ${}_{\vert i}$ the spatial derivative with respect to $X_i$, by an overdot the partial time derivative, $ \epsilon_{ijk} $ being the Levi--Civita symbol. 
The volume deformation of fluid elements is described \footnote{It is initially equal to ${J(\mathbf{X},\initial t)}=1$ since, by definition, Eulerian coordinates initially coincide with Lagrangian ones.} by $J(\mathbf{X},t) $.
To obtain the Lagrangian formulation of Newtonian gravity, we define the Eulerian fields in terms of $\mathbf{f}$ and its derivatives.
The position, velocity, acceleration, density, vorticity, {\it etc.} can be given as functionals of the trajectory function:
\begin{equation}
\mathbf{x}:= \mathbf{f}(\mathbf{X},t) \;;\; \mathbf{v}:= \dot{\mathbf{f}}(\mathbf{X},t) \; ;\; \mathbf{a}:= \ddot{\mathbf{f}}(\mathbf{X},t) \;;\;
\varrho(\mathbf{X},t)=\frac{\mathring{\varrho}}{J(\mathbf{X},t)} \;;\; \boldsymbol{\omega} = \frac{\mathring{\boldsymbol{\omega}}\cdot \boldsymbol{\nabla}_{\bf 0} \mathbf{f}}{J(\mathbf{X},t)}\;\,; \;etc. \, ,
\label{identi}
\end{equation}
where $\mathring{\varrho}$ and $\mathring{\boldsymbol{\omega}}$ stand for the initial values of the density and vorticity fields, and
$\boldsymbol{\nabla}_{\bf 0}$ for the nabla operator with respect to the Lagrangian coordinates.\par
Contrary to the Eulerian description, a single variable describes the gravitational dynamics: $\mathbf{f}$.
In the Lagrangian approach, the Eulerian position $\mathbf{x}=\mathbf{f}(\mathbf{X},t)$ is no longer an independent variable. The independent variables are now $(\mathbf{X},t)$. \par
According to the equivalence of inertial and gravitational masses, we can express the field strength in terms of the acceleration  $\mathbf{g}=\mathbf{a}=\ddot{\mathbf{f}}$; also, we have to transform the Eulerian spatial derivatives in the field equations to Lagrangian coordinates:
\begin{eqnarray}
\label{LNS1}\delta_{ab} \ddot{f}^{a}_{\;\,\vert [i}\,  {f}^{b}_{\;\,\vert j]} = 0\quad;\quad \frac{1}{2} {\epsilon}_{abc} {\epsilon}^{ikl}\ddot{f}^{a}_{\;\,\vert i}  \, f^b_{\;\,\vert k} \, f^c_{\;\,\vert l} = \Lambda J - 4 \pi G \mathring{\varrho} \;.
\end{eqnarray}
This system of equations is the {\em Lagrange-Newton system}.\par

In the Lagrangian perturbative approach, the trajectory field is decomposed into the sum of a homogeneous deformation plus an inhomogeneous deformation field:
\begin{equation}
\mathbf{f}(\mathbf{X},t) = a(t) \left( \mathbf{X} +\mathbf{P}(\mathbf{X},t)\right)\:\:.
\end{equation}
The inhomogeneous perturbations can be decomposed into different orders $n$.
Due to the facts that the perturbed flow is a function of the Lagrangian coordinates (following the fluid flow) and that the trajectory function is the only perturbed variable,
deviations from the homogeneous flow may be small while the Eulerian fields, evaluated along the perturbed flow, can experience large changes. 
Another way to phrase this aspect is to say that the Lagrangian description of structure formation intrinsically contains nonlinear Eulerian terms.
In the frame of Newtonian gravity, the Lagrangian approach has proved to be more powerful than the Eulerian one due to this reason.\par In the next section we will present the Lagrangian formulation of the Einstein equations. Since in general relativity matter is coupled to the geometry of spacetime, we give up the notion of absolute space and time, and the gradient of the vector function $\mathbf{d} f^a$ that describes the embedding into the absolute vector space has to be replaced by nonexact one--form fields, the Cartan coframes. As we will discuss, this relativistic formulation exhibits interesting formal similarities with the Newtonian theory. This can be exploited to build relativistic solutions to any order.

\subsection{Relativistic intrinsic Lagrangian theory}

Spacetime can be foliated into flow--orthogonal hypersurfaces of constant synchronous time for the fluid model {\it irrotational dust}. 
We here consider a Lagrangian splitting of the Einstein equations according to the $3+1$ formalism. \par
As in the Newtonian case, we consider a local Lagrangian coordinate basis $\{\mathbf{d} X^i\}$, which spans the local cotangent spaces on the manifold.
The dynamics of the fluid is no longer described by the gradient of the Newtonian trajectory function $\mathbf{d} f^a = {f}^a_{\ \vert i}\mathbf{d} X^i$, but by its  relativistic counterpart:  the nonintegrable Cartan coframes $\boldsymbol{\eta}^a= \eta^a_{\ i} \mathbf{d} X^i$, where $a=\{1,2,3\}$ count the one--forms. These coframes encode both the geometry of spacetime and the dynamics of the fluid. The $4$--dimensional metric can be decomposed according to 
\begin{equation}
\label{metric}
^{4}\mathbf{g} = -\mathbf{d}t \otimes \mathbf{d}t + ^{3}\mathbf{g}\quad \mbox{where}\quad ^{3}\mathbf{g} = G_{ab} \boldsymbol{\eta}^a \otimes  \boldsymbol{\eta}^b\:\:\Rightarrow g_{ij} = G_{ab} \eta^a_{\ i}\eta^b_{\ j}\:\:;
\end{equation}
$G_{ab}$ is the Gramian matrix of the coframes $\boldsymbol{\eta}^a$. This constant matrix can be chosen such that it encodes the information on the initial deviation fields.
\par
Inserting the metric expression \eqref{metric} into the Einstein equations, we  obtain the following system of equations, called the {\em Lagrange--Einstein system}:\cite{rza1}
\begin{align}
\label{form_symcoeff}&G_{ab} \,\ddot{\eta}^a_{[i} \eta^b_{\ j]} = 0 \;; \\
\label{form_eomcoeff}&\frac{1}{2 J} \epsilon_{abc} \epsilon^{ikl} \left( \dot{\eta}^a_{\ j} \eta^b_{\ k} \eta^c_{\ l} \right) \dot{} = -\CR^i_{\ j} + \left( 4 \pi G \varrho + \Lambda \right) \delta^i_{\ j}\;; \\
\label{form_hamcoeff} &\frac{1}{2J}\epsilon_{abc} \epsilon^{mjk} \dot{\eta}^a_{\ m} \dot{\eta}^b_{\ j} \eta^c_{\ k} = - \frac{\CR}{2}+ \left( 8\pi G \varrho +  \Lambda  \right) \;; \\
\label{form_momcoeff}&\left(\epsilon_{abc} \epsilon^{ikl} \dot{\eta}^a_{\ j} \eta^b_{\ k} \eta^c_{\ l} \right)_{||i} = \left(\epsilon_{abc} \epsilon^{ikl} \dot{\eta}^a_{\ i} \eta^b_{\ k} \eta^c_{\ l} \right)_{|j} \;.
\end{align}
The system $\lbrace\eqref{form_symcoeff}-\eqref{form_momcoeff}\rbrace$ consists of $13$ equations, comprising $9$ evolution equations for the $9$ coefficient functions of the $3$ coframe fields and $4$ evolution equations that originate from the spatial $3+1$ constraints.
The  double vertical slash denotes the covariant spatial derivative with respect to the $3$--metric and the spatial connection is assumed to be symmetric.
We can eliminate the intrinsic curvature $\CR^i_{\ j}$ by inserting the trace of the second equation into the third to obtain the Raychaudhuri equation:
\begin{eqnarray}
\label{Raycoeff}
\epsilon_{abc}\epsilon^{ik\ell} \ddot{\eta}^a_{\ i} \eta^b_{\ k} \eta^c_{\ \ell}   = \Lambda J - 4 \pi G \mathring{\varrho} \;.
\end{eqnarray}
As we will demonstrate below, \eqref{form_symcoeff} and \eqref{Raycoeff} can be expressed as identities on the electric part of the Weyl tensor. This forms the gravitoelectric part of the Einstein equations being formally similar to \eqref{LNS1}, where $f^a_{\ \vert i}$ is replaced by ${\eta}^a_{\ i}$: the gravitoelectric perturbative solutions can therefore be built from the Newtonian ones. \cite{rza3} 

\section{The Maxwell--Weyl equations, the relativistic generalization of the Newtonian perturbative solutions, and gravitational waves}

The Weyl tensor is the traceless part of the $4$--Riemann curvature tensor \cite{Ellis:1998ct} representing the part of the gravitational field that is not directly coupled to the matter content of the Universe. It can be split in an irreducible way into electric and magnetic parts, corresponding to two symmetric and traceless tensors, each of them containing $5$ independent components. Contracting the Bianchi identities, which link the covariant derivatives of the Weyl tensor to the ones of the Ricci curvature tensor, we can formulate general relativity in terms of the electric and magnetic parts of the Weyl curvature tensor.
In the restframe of the irrotational dust fluid, with $4-$velocity ${u}^{\mu}=(1,0,0,0)$, we obtain the Maxwell--Weyl equations in the present framework.\cite{rza1}
We only write here the spatial components of the electric part,
\begin{equation}
E^i_{\ j}  := -\dot{\Theta} ^i_{\  j}  -\Theta^{i}_{\ k}  \Theta^{k}_{\ j}  -\frac13  \left( 4 \pi G \varrho - \Lambda \right) \delta^i_{\ j} \:\:,
\end{equation}
which can be expressed through the coordinate components of Cartan coframes:
\begin{equation}
E^i_{\ j} = -\frac{1}{2J} \epsilon_{abc} \epsilon^{ikl} \ddot{\eta}^a_{\ j} \eta^b_{\ k} \eta^c_{\ l} + \frac{1}{3}\left( \Lambda  - 4 \pi G \frac{\mathring{\varrho}}{J} \right)\delta^i_{\ j} \;\;;\;\; J = \det (\eta^a_{\ i} ) \label{ge}\ .
\end{equation}
The electric part of the Weyl tensor is thus (up to a sign) the relativistic counterpart of the Newtonian tidal tensor 
${\mathcal E}^i_{\ j} = \partial g^i / \partial x^{j} - 1/3 \delta^i_{\; j} \boldsymbol{\nabla} \cdot {\bf g}$. The gravitoelectric part of the Einstein equations,
\begin{equation}
E^k_{\ k}  = 0 \quad \Leftrightarrow \quad \eqref{Raycoeff} \quad ; \quad E_{[ij]}  = 0 \quad \Leftrightarrow \quad \eqref{form_symcoeff}\:\:, \label{EElectric}
\end{equation}
formally corresponds to the (closed) {\em Lagrange--Newton system}, 
if we replace the nonintegrable coefficients of the Cartan coframes by integrable ones,  ${\eta}^{a}_{\ i} \rightarrow f^{a }_{\ \vert i}$, keeping the speed of light $c$ finite.
This mathematical restriction can be reversed in order to obtain relativistic solutions, denoted by ${}^{E} \mathbf{P}^a$, from the Newtonian ones
through execution of the following transformation rule: 
\begin{equation}
\mathbf{d} f^i = a(t) \left( \mathbf{d} X^i +\mathbf{d} P^i  \right) \rightarrow \boldsymbol{\eta}^a = a(t) \left( \mathbf{d} X^a +{}^E\mathbf{P}^a  \right)\:.
\end{equation}
(We refer the reader to the paper \cite{rza3} for explicit solutions.)
In the next section, we will employ the Hodge decomposition in the frame of a global description. Assuming a closed topology for the spatial sections, we will  investigate the physical content of the complementary part ${}^H\mathbf{P}^a := \mathbf{P}^a - {}^E\mathbf{P}^a$ in the general perturbation, and
we will then discuss the link between these two solutions and the magnetic part of the Weyl tensor. 
This also sheds a new light on the relation between topology and geometry.

\section{From topology to gravitational waves}

We recall that the perturbation fields are defined intrinsically on the perturbed spatial sections. 
The powerful Hodge decomposition of Cartan coframes will provide new results on general first--order perturbations and gravitational waves.

We first note that the gravitomagnetic part that is generated by the first--order gravitoelectric solution, $H_{ij}(^{E} \mathbf{P}^a)$, turns out to be harmonic:\cite{rza4}
\begin{equation}
\Delta_{\bf 0} H_{ij}(^{E} \mathbf{P}^a)=0\:\:;
\label{harm}
\end{equation}
$\Delta_{\bf 0}$ denotes the Laplacian operator with respect to local (Lagrangian) coordinates.\par
The spatial topology determines the boundary conditions needed to solve this equation. Thurston's geometrization conjecture presented in \cite{thurston} (now proved by Grigori Perelman) asserts that every closed $3$--manifold can be decomposed into a connected sum  of $3$--manifolds modeled after the $8$ model geometries listed by William Thurston. 
This asserts that any closed {\em and simply--connected} $3$--manifold is homeomorphic to the hypersphere $\mathbb{S}^3$. Moreover, as a consequence of parallelizability, the Cartan coframes are defined on the whole manifold continuously while the coordinate charts undergo singularities. In what follows, we will assume $\mathbb{S}^3$ for the topology of spatial sections as an example.
 
The Hodge decomposition\cite{hodge}\footnote{For the presentation of the context around the Hodge theorem, from which the Hodge decomposition has been obtained, the reader may refer to \cite{Lee} and references therein.} decomposes $p$--forms into an exact part, a coexact part and a harmonic part $\mathbf{h}^a = h^a_{\ i} \,\mathbf{d}X^i$. Harmonicity is defined with respect to the Laplace--De Rham operator $\boldsymbol{\Delta}^{dR}$ that encodes the geometry of the manifold considered. On a coordinate basis $\mathbf{d} X^i$, according to the formula of Weitzenb\"ock,
\begin{equation}
0= \left( \boldsymbol{\Delta}^{dR}\ \mathbf{h}^{a} \right)_i = \left( \boldsymbol{\Delta}^B \,\mathbf{h}^a \right)_i +  {h}^a_{\ k} \: \mathcal{R}^{k}_{\ i}\:\:;\;\; \boldsymbol{\Delta}^B \,\textbf{h}^a : = -  \textbf{h}^{a}{}^{\parallel k}_{\ \ \parallel k}  \;\;,
\end{equation}
where $\boldsymbol{\Delta}^B$ is the Laplace--Beltrami operator.
The Hurewicz theorem implies that the dimension of the harmonic space of one--forms on a closed oriented manifold is equal to the dimension of the first fundamental group (also called closed path group). \cite{Brezis,deRham,Dieu} For 
simply--connected manifolds the dimension of this group is null, e.g. for the $\mathbb{S}^3$ topology. The Laplace equation \eqref{harm} thus implies, for an $\mathbb{S}^3$ topology, that $H_{ij}(^{E} \mathbf{P}^a)=0$. Furthermore, the Cartan coframes, which are one--forms, can be decomposed on $\mathbb{S}^3$  as
$\boldsymbol{\eta}^a =  \mathbf{d}\alpha^a+ \boldsymbol{\delta}  \boldsymbol{\beta}^a$,
where $\alpha^a$ is a scalar, $ \boldsymbol{\beta}^a $ a two--form,
$\mathbf{d}$ the exterior derivative and $ \boldsymbol{\delta} = \star \mathbf{d} \star$ its dual, $\star$ the Hodge star operator. \par
The magnetic part is linked at first--order to the Cartan structure coefficients:\cite{rza4} $C^a_{\ bc}$, $\mathbf{d}\boldsymbol{\eta}^a = - C^a_{\ bc} \boldsymbol{\eta}^b \wedge \boldsymbol{\eta}^c$ :  $H^{i}_{\ j} = \frac{1}{2} \epsilon_{j}{}^{kl} \dot{C}^i_{\ kl}\,$.
Therefore, the Cartan structure coefficients calculated from $^{E} \mathbf{P}^a$ are constant in time. Since they are initially zero ($\boldsymbol{\eta}^a (t_i) = \mathbf{d} X^a )$, $^{E} \mathbf{P}^a$ represents the integrable part  $\mathbf{d}\alpha^a$ of the full perturbation. Deviations from Euclidean space, at first order, are thus fully contained in the complementary part
$^{H}\mathbf{P}^a$ that turns out to be trace--free. Gravitational waves, as a non--Newtonian effect, are contained in ${}^{H}\mathbf{P}^a$, the only curvature--generating part of the perturbation. These results can be extended to the quotients of $\mathbb{S}^3$. 

\section{Conclusion and outlook}

We discussed how the gravitoelectric part of relativistic perturbations can be constructed from the Newtonian theory.
At first order strong results for the general perturbation fields can be established by global considerations: for an $\mathbb{S}^3$ topology of spatial sections the gravitoelectric part is integrable, while the nonintegrable part coincides with the gravitomagnetic part that is linked to the gravitational waves.
Formulating the Einstein equations in terms of the Hodge decomposition of Cartan coframes may be of great use for upcoming research. Investigating the extension of these results to higher orders and other topologies will then become easier.

\section*{Acknowledgments}

Thanks to Alexandre Alles, L\'eo Brunswic, Mauro Carfora, Martin Kerscher, Pierre Mourier, Matthias Ostermann, 
Jan Ostrowski, Boud Roukema, Xavier Roy, Herbert Wagner and Alexander Wiegand for constructive discussions. 
This work was conducted within the ``Lyon Institute of Origins'' under grant ANR--10--LABX--66. 
We acknowledge support by the \'Ecole Doctorale PHAST Lyon, and by 
the French--Bavarian Cooperation Center, BFHZ Munich \url{http://www.bayern-france.org/}.


\begin{thebibliography}{100}

\bibitem{rza1}
T.~Buchert and M.~Ostermann,
{\em Phys. Rev. D} \textbf{86}, 023520 (2012). 

\bibitem{rza3}
A.~Alles, T.~Buchert, F.~Al Roumi and A.~Wiegand, {\em Phys. Rev. D} \textbf{92}, 023512 (2015). 

\bibitem{ehlersbuchert}
J.~Ehlers and T.~Buchert,
{\em Gen. Rel. Gravit.} \textbf{29}, 733 (1997). 

\bibitem{Ellis:1998ct}
G.~F.~R.~Ellis and H.~van Elst,
{\em NATO Sci. Ser. C} \textbf{541}, 1 (1999). 
    
\bibitem{rza4}
T.~Buchert, F.~Al Roumi and A.~Wiegand, {\em in preparation}.

\bibitem{thurston}
W.~P.~Thurston,
The Geometry and Topology of Three--Manifolds, Vol. 1,
Princeton University Press (1997).

\bibitem{hodge}
W.~V.~D.~Hodge, The Theory and Applications of Harmonic Integrals, Cambridge University Press (1941).

 \bibitem{Brezis}
H. Brezis,
Analyse fonctionnelle: th\'eorie et applications,
Dunod, Paris (2005). 

\bibitem{deRham}
G. De Rham,
Sur la th\'eorie des formes diff\'erentielles harmoniques,
{\em Annales de l'Universit\'e de Grenoble} \textbf{22}, 135--152 (1946).

\bibitem{Dieu}
J. Dieudonn\'e,
El\'ements d'analyse, Tome 1, Gauthiers Villars, Paris, (1972).

\bibitem{Lee}
J.M. Lee, Introduction to Smooth Manifolds, 2nd edition, {\em Graduate Texts in Mathematics} \textbf{218}, Springer, New York (2013).



\end{thebibliography}
\end{document}